# Engineering Defect-Phonon Interactions Through Heterophase-Interfaces in Silicon Carbide Membranes

*Kirlie Iulius Figuera Michal[1,†], Jin Hee Lee[1,†], Keiju Sato[2], Takuji Maekawa[2], Je-Hyung Kim[1*]*

[1]Department of Physics, Ulsan National Institute of Science and Technology (UNIST), Ulsan 44919, Republic of Korea.

[2]ROHM Research & Development Center, ROHM Co., Ltd., Kyoto 615-8585 Japan

## ABSTRACT

Point defects in wide-bandgap semiconductors offer spin and photonic qubits in a solid-state platform, making them important building blocks for quantum information processing, communication, and sensing. While these systems have the strong advantage of room-temperature operation, intrinsic electron-phonon interaction induces broad phonon-sideband emission and weak zero-phonon-line transitions, limiting efficient spin–photon interfaces and scalable photon-mediated interaction. Here, we introduce a controlled heterophase interface based on the remote epitaxy technique as a crystal heterogeneity-engineering strategy. While stacking faults are treated as imperfections to be eliminated, our results instead show that crystal-phase interfaces can provide an additional degree of freedom to engineer defect-phonon interaction beyond the intrinsic properties of a single crystal. In comparison with single-phase 4H-silicon carbide (SiC) membranes, Vsi in 3C/4H heterointerface SiC exhibits a drastic enhancement of zero-phonon optical transitions, with a substantially narrower zero-phonon linewidth of 4.71 meV even at room temperature. These results establish heterophase-interface engineering as an effective route for tailoring defect–phonon interactions and realizing bright room-temperature quantum emitters with enhanced zero-phonon optical transitions.

## INTRODUCTION

Point-defect color centers in wide-bandgap semiconductors have emerged as a promising source of single photons and optically addressable spin qubits for a range of quantum technologies, such as quantum communication, quantum sensing, and quantum information processing.[1-4] Compared with other quantum emitters, these color centers offer key advantages, including stable optical emission, compatibility with semiconductor fabrication infrastructure, and the potential for room-temperature operation.[1, 5]

Following early breakthroughs based on nitrogen-vacancy (NV) centers in diamond, significant progress has been made in enhancing photon extraction efficiency, extending spin coherence times, and realizing spin–photon interfaces.[6-8] Despite their outstanding spin coherence properties, NV centers exhibit broad emission spectra, with only a small fraction of photons emitted into the zero-phonon line (ZPL) emission.[9, 10] However, high-fidelity spin-photon entanglement and efficient photon-mediated interconnection require optical emission occurring preferentially in the ZPL emission.[11, 12] Achieving a strong ZPL emission remains a major challenge because optical transitions in solid-state defect systems commonly couple to lattice vibrations, redistributing emission into broad phonon sidebands and reducing spectral purity as temperatures increases.[13, 14] Consequently, controlling these electron–phonon interactions to maintain a strong ZPL remains an outstanding hurdle for establishing scalable quantum system.[15, 16]

Efforts to achieve strong ZPL emission have traditionally relied on exploiting the intrinsic crystal symmetry and electronic structure of specific defect systems. For example, in diamond, group-IV vacancy centers such as the silicon-vacancy ($SiV^-$) and tin-vacancy ($SnV^-$) exhibit strong ZPL emission, in part because their inversion-symmetric structure reduces sensitivity to local electric-field fluctuations and improves optical spectral stability.[17-20] However, such favorable properties are intrinsically governed by the crystal symmetry and electronic structure of a given host–defect system, posing a challenge to generalization and deterministic engineering.

Here, we investigate an engineering route for point defects-phonon interactions by introducing stacking faults and forming crystalline heterophase interfaces within the materials. In conventional materials processing, stacking faults are typically regarded as undesirable imperfections and are actively minimized to preserve crystal uniformity. However, these unique structural planar defects can instead be harnessed as a controllable parameter for engineering electronic and vibrational properties at the nanoscale.

Silicon carbide (SiC) provides a particularly suitable platform for this approach because its more than 250 known polytypes arise from the small energetic differences between alternative stacking sequences of Si–C bilayers, giving SiC one of the richest displays of polytypism among solid-state systems.[21] Furthermore, SiC hosts a variety of point defects,[22] including the silicon vacancy (Vsi), which exhibits long spin coherence times, optically addressable spin states, and robust spin properties even at room temperature.[23, 24]

Since different SiC polytypes possess distinct lattice and electronic structures, heterophase interfaces can create unique interfacial environments that modify local symmetry and vibrational characteristics, thereby influencing electron–phonon interactions and defect optical transitions.[25, 26] Therefore, such modification of local interfacial environments offers a promising pathway to engineering intrinsic phonon interaction nearby point defects. A representative example was recently demonstrated in SiC nanowires containing stacking faults, where point defects exhibited exceptionally strong room-temperature zero-phonon transitions and high Debye–Waller factors.[27] However, because these defect–interface configurations emerge stochastically during growth, their optical properties remain inherently difficult to reproduce and control.

In this work, we demonstrate that the formation of heterophase-interface in SiC provides a direct and effective pathway to suppress electron–phonon coupling and enhance the zero-phonon transition of Vsi centers. Using remote epitaxy on graphene, we realize transferable SiC membranes in which the stacking sequence can be controlled through the underlying graphene interface. This approach enables the formation of both single-crystalline 4H-SiC membranes and polytype-mixed structures containing 4H–3C stacking faults. By directly comparing these systems, we reveal that the presence of stacking faults leads to a pronounced enhancement of the ZPL emission from Vsi centers, indicating a substantial modification of electron–phonon coupling. Given the wide availability of polytypism in other quantum materials, our work establishes heterophase-interface engineering as an efficient strategy for overcoming intrinsic phonon limitations and tailoring the optical properties of solid-state quantum emitters.

## RESULTS AND DISCUSSION

A high-crystal-quality epitaxial thin film of SiC can be grown on a SiC substrate using the chemical vapor deposition (CVD) method. We introduce a remote epitaxy technique, in which a graphene interlayer is first prepared on a single-crystalline SiC substrate, followed by epitaxial growth of a SiC membrane that is epitaxially aligned with the underlying substrate through the atomically thin graphene interlayer,[28] as schematically illustrated in Figure 1a. A Ni stressor layer and thermal-release tape are subsequently used to mechanically exfoliate the grown membrane from the parent SiC substrate; further growth and exfoliation details are provided in the Methods.

Importantly, the graphene interlayer directly affects the lattice arrangement of the overgrown layer.[29, 30] Because SiC growth proceeds via a step-flow mechanism,[31] this growth geometry enables intentional control over the resulting crystal phase: growth initiates at the step edges of the substrate, and the local graphene thickness determines whether the substrate's crystal information is transmitted to the overgrown layer. When the steps are covered by sufficiently thin graphene (*i.e.,* a few layers), the substrate's electrostatic potential penetrates the interlayer, transmitting the underlying stacking information and yielding 4H-SiC growth.[32] In contrast, when the steps are covered by thicker graphene that screens this potential, 3C-SiC is expected to nucleate instead. By alternating the graphene thickness at the step edges, a 4H-SiC/3C-SiC heterostructure can therefore be intentionally fabricated.

Using this mechanism, we introduce a heterophase-interface engineering strategy that intentionally creates a 3C/4H-SiC heterostructure containing polytype mixing, resulting in the formation of heterophase-interface-associated-point defects. This approach is expected to modify the interfacial growth environment, providing opportunities to control nucleation, strain relaxation, and crystal-phase evolution in the epilayer.[33]

To verify that this graphene-thickness-controlled growth indeed yields the intended structural outcomes, we characterized the exfoliated SiC membranes by cross-sectional transmission electron microscopy (TEM). Figure 1b and 1c compare TEM images of membranes grown on a monolayer graphene interlayer and a multilayer graphene interlayer, respectively. The membrane grown on the monolayer interlayer preserves a single-crystalline 4H-SiC structure, consistent with the underlying substrate, as shown in Figure 1b, whereas the membrane grown on the multilayer interlayer reveals heterophase interfaces with coexisting 4H- and 3C-SiC domains embedded within the membrane, as shown in Figure 1c.

In addition to the cross-sectional TEM, we conducted confocal Raman spectroscopy (Alpha300R, WITec) with 532 nm excitation to spatially map the lateral distribution of crystal phases across the membrane. Figure 2a, d shows optical microscope images of transferred flakes of a SiC membrane with 4H single phase and 3C/4H heterophase. The samples were spatially scanned at different Raman shift frequencies, corresponding to 4H- and 3C-SiC. The resulting Raman intensity maps are presented in Figures 2b and 2c for the single-phase membrane. Across the transferred membrane region, a strong and spatially continuous 4H-SiC Raman response is observed, whereas the corresponding 3C-SiC response remains negligible in the 4H single-phase membrane. These observations indicate that the membrane preserves a phase-pure 4H crystalline structure following remote epitaxial growth and membrane transfer.

We next examined a second transferred membrane region selected for heterophase-interface characterization. In contrast to the phase-pure membrane, the Raman confocal maps shown in Figures 2e and 2f reveal spatial coexistence of both 4H-SiC and 3C-SiC Raman signals. indicating spatial distribution of heterophase interface over a transferred SiC membrane.

To further verify the different configuration of crystal phases of the SiC, Raman spectra acquired from transferred membranes are also compared with conventional bulk 4H SiC and 3C SiC on insulator samples. The resulting spectra are summarized in Figure 2g. The upper two spectra correspond to bulk 4H-SiC and bulk 3C-SiC references, while the lower spectra were obtained from the single-phase and heterophase membranes, respectively. Each gray guidelines indicate the characteristic transverse optical phonon modes of 4H-SiC (~776 $cm^{-1}$) and 3C-SiC (~796 $cm^{-1}$).[34] The single-phase membrane reproduces the Raman response of bulk 4H-SiC, whereas the heterophase membrane simultaneously exhibits Raman features associated with both 4H and 3C polytypes, confirming successful formation of 4H/3C polytype mixing in a thin SiC membrane.

Combined confocal micro-photoluminescence (μ-PL) measurements enable the identification of optically active point defects and provide a direct correlation between local crystal phase structure with points defect's optical transitions. Confocal μ-PL was performed at room temperature under continuous-wave off-resonant excitation at 780 nm with an incident power of 1 mW. We first examined single-phase 4H-SiC membrane regions. The confocal PL map in Figure 3a reveals multiple localized emission sites distributed across the membrane, with several bright spots reaching count rates approaching 200 kcps. In Figure 3b, these point defects exhibit broad near-infrared emission as a result of phonon interaction similar to that observed from Vsi in the bulk 4H-SiC substrate (Figure S1) and not clearly resolved narrow zero-phonon transition is observed. We next performed the measurement for the 3C/4H heterophase SiC membrane. The heterophase membrane also exhibits well-isolated, bright point defects (Figure 3c). However, their spectral feature differs markedly from those of the single-phase membranes. A representative emitter exhibits a narrow room-temperature emission line centered at 873.8 nm (1.419 eV), as shown in Figure 3d. This pronounced suppression of phonon-sideband emission observed in the heterophase membrane provides clear evidence that the 3C/4H heterophase interface substantially modifies the defect–phonon interaction. The spectral positions of the heterophase-interface-associated defects vary from emitter to emitter; however, they consistently exhibit unusually narrow room-temperature zero-phonon transitions (Figure S2).

To evaluate the quantum nature of this emitter, second-order autocorrelation measurements were performed with point defects with a strong ZPL at room temperature. Figure 3d inset shows a clear antibunching dip with a raw $g^{(2)}(0) = 0.25$ without any background correction, confirming room-temperature single-photon emission. Therefore, these measurements demonstrate that point defects with 3C/4H heterophase interfaces support bright single-photon emission at narrow zero-phonon optical transitions at room-temperature.

We further investigate the spectral characteristics of these defects within a 3C/4H heterophase membranes with temperature-dependent μ-PL spectroscopy. At 4 K, the PL spectrum exhibits narrow peaks centered near 865 nm, as shown in Figure 4a. Gaussian fitting yields linewidths of approximately 0.042 nm (~70 μeV) for the individual peaks, close to the spectral resolution limit of the measurement system.

Second-order correlation measurements, $g^{(2)}(\tau)$, were performed at 4 K using a broadband spectral filter covering both emission peaks. As shown in Figure 4b, the autocorrelation function exhibits a pronounced antibunching dip with a raw $g^{(2)}(0) = 0.008$ without background correction, confirming single-photon emission from a single quantum emitter. To further verify the origin of the doublet, autocorrelation measurements were repeated using narrower spectral filters that selectively isolated each peak (Figure S3). Both measurements yielded an identical uncorrected value of $g^{(2)}(0) = 0.008$, demonstrating that the two emission peaks originate from the same single emitter rather than from two independent defects. The observed doublet is therefore likely associated with different charge states of the same defect or spectral hopping induced by fluctuations in the local charge environment.

The temperature-dependent PL spectra are shown in Figure 4c. As temperature increases, thermal linewidth broadens occurs, but the ZPL is clearly observable up to room temperature. For comparison, the point defect in a single-phase 4H-SiC membrane was also characterized with temperature. At a cryogenic temperature of 4 K, the emitter also reveals narrow ZPLs. However, phonon-sideband emission progressively dominates the spectrum above 100 K, making the ZPL difficult to resolve.

The strong zero-phonon transition in our heterophase membranes enables us to directly trace the thermal evolution of the ZPL in SiC across the full temperature range (4–300 K), as shown in Figure 4d. Below 50 K, both the ZPL linewidth (FWHM, obtained from Voigt profile fits) and peak energy remain nearly constant, indicating an inhomogeneous limit dominated by temperature-independent dephasing mechanisms. Above 50 K, thermal excitation leads to a simultaneous broadening of the linewidth ($\Gamma(T) = \Gamma_0 + aT^3$) and a spectral redshift of the emission energy ($E(T) = E_0 + bT^2$) with temperature ($T$). This combination of $T^3$ dependent broadening and $T^2$ dependent energy shift reflects soft-mode dynamics, a behavior previously reported for visible color centers in SiC[35] and other wide-bandgap quantum emitters.[36, 37] Given the narrow and well-resolved ZPL in the heterophase membranes, the temperature-dependent optical response of defects enables optical thermometry.[38] Near room temperature, the emitters show a ZPL linewidth of 2.89 nm and temperature sensitivity of 0.058 nm/mK. Such approaches have previously been demonstrated primarily for room-temperature emitters such as SiV centers in diamond.[39] However, SiV is known to have low quantum yield at high temperature.[40] Therefore, the bright zero-phonon transition over a wide temperature range could provide a practical route toward sensitive all-optical thermometry.

The observed room-temperature linewidth of 4.7 meV is comparable to those reported for inversion-symmetric group-IV color centers in diamond, such as $SiV^{-}$[41] and $SnV^{-}$[42]. More importantly, the optical characteristics of the heterophase-interface-associated defects in SiC closely resemble those of many room-temperature visible quantum emitters reported in GaN[43] and hBN[44], including bright zero-phonon emission and large emitter-to-emitter spectral inhomogeneity. Despite extensive studies in both material systems, the microscopic origin of these emitters remains under active debate. GaN typically contains a high density of extended defects, such as threading dislocations and basal-plane stacking faults, while the layered structure of hBN naturally gives rise to stacking variations and other structural imperfections.[45] Our results demonstrate that the presence of crystal-phase interfaces can profoundly modify the optical properties of point defects, giving rise to narrow zero-phonon transitions accompanied by substantial spectral inhomogeneity. These observations suggest that these types of spectrally narrow, room-temperature emitters with spectral inhomogeneity may originate from defect complexes or point defects coupled to local crystal-phase interfaces, rather than from a single, well-defined point-defect species.

## CONCLUSION

In summary, we have demonstrated that heterophase-interface engineering provides a direct and effective pathway to suppress point defect–phonon coupling and enhance zero-phonon optical transitions. By introducing 3C/4H heterophase interfaces in remotely epitaxial SiC membranes, we realized bright, spectrally narrow quantum emitters exhibiting near-unity single-photon purity at low temperature and persistent, well-resolved ZPL emissions up to room temperature. This is in contrast to phase-pure 4H-SiC, where phonon-sideband emission dominates with temperature. Polytypism and stacking faults have traditionally been viewed as undesirable structural defects that degrade crystal quality and device performance, but our work instead demonstrates that crystal-phase interfaces can serve as functional elements for controlling defect optical properties, introducing a new degree of freedom for engineering solid-state quantum emitters beyond the intrinsic symmetry and electronic structure of the defect.

Crystal-phase interfaces, stacking faults, and related structural heterogeneities are ubiquitous in many wide-bandgap semiconductors, so this strategy is expected to be broadly applicable to a wide range of defect-hosting materials beyond SiC, including hexagonal diamond phase[46]. Together with the scalability offered by remote epitaxy and the compatibility of SiC with integrated photonic platforms, crystal-phase engineering provides a promising route toward developing solid-state quantum emitters with tailored optical properties for quantum photonics, quantum sensing, and quantum information technologies.

## Methods

### Sample Information

First, epitaxial graphene was synthesized on a single-crystal 4H-SiC (0001) substrate, 4 degrees off-axis toward the [11-20] direction, via thermal decomposition of the substrate.[47] Following the epitaxial graphene growth, a SiC epitaxial layer was grown on top of the graphene/SiC template by CVD. To preserve the underlying graphene layer and enable successful epitaxial growth compatible with remote epitaxy, the CVD process was conducted under Silane, Propane and Ar flow conditions using optimized growth parameters to mitigate the etching of the graphene release layer. Finally, a Ni stressor layer and thermal-release tape were introduced to mechanically exfoliate the overgrown SiC membrane from the parent SiC substrate.[48]

### Optical Characterization Setup

Room-temperature optical characterization was performed using a home-built confocal microscope equipped with a high-numerical-aperture objective lens (NA=0.95) and a 780 nm continuous-wave (CW) laser. Confocal PL maps were acquired using a piezoelectric XYZ scanning stage. The PL emission was isolated using a combination of a dichroic mirror, a long-pass filter, and a band-pass filter before being collected into a 50:50 single-mode fiber beam splitter. The signal was subsequently routed to either a spectrometer or an avalanche photodiode.

For low-temperature measurements, the sample was housed in a closed-cycle cryostat (Montana Instruments) maintained at 4 K. A long-working-distance objective lens (NA=0.7) was utilized for both excitation (780 nm CW laser) and collection. The emission was coupled into a single-mode fiber using an aspheric lens and routed to either the spectrometer or a superconducting nanowire single-photon detector. For Hanbury Brown and Twiss interferometry measurements, the emission was spectrally isolated using either a 50 nm free-space band-pass filter or a narrow fiber band-pass filter with a 0.15 nm transmission window, before being split into two distinct detection paths using a fiber beam splitter.

## Author Information

**Corresponding Authors**

Correspondence should be addressed to J.-H.K.

**Je-Hyung Kim**

Department of Physics, Ulsan National Institute of Science and Technology (UNIST), Ulsan 44919, Republic of Korea

orcid.org/0000-0002-6894-9285;

*Email: jehyungkim@unist.ac.kr

**Author Contributions**

†K.I.F.M and J.H.L. contributed equally to this work.

T.M. and J.-H. K conceived the idea. K. S. and T. M. K were responsible for sample growth, membrane preparation, and TEM measurement. K.I.F.M and J.H.L. performed micro-Raman, PL, and correlation measurements. T.M. and J.-H. K. supervised the research. All authors discussed the results and commented on the manuscript.

## Acknowledgement

This work was supported by the National Research Foundation (RS-2024-00438839; RS-2025-02317602) and the Institute for Information & Communications Technology Planning & Evaluation (IITP) Grant (RS-2025-25464832) of Korea. The authors thank the Research Fund (1.250007.01) of UNIST (Ulsan National Institute of Science & Technology) and UNIST Central Research Facilities (UCRF) QuantumNanoFab supported by IITP (RS-2024-00401037). This work was partly implemented under a joint research project of Tsukuba Power Electrics Constellations (TPEC).

## Supporting Information

**FIGURES**

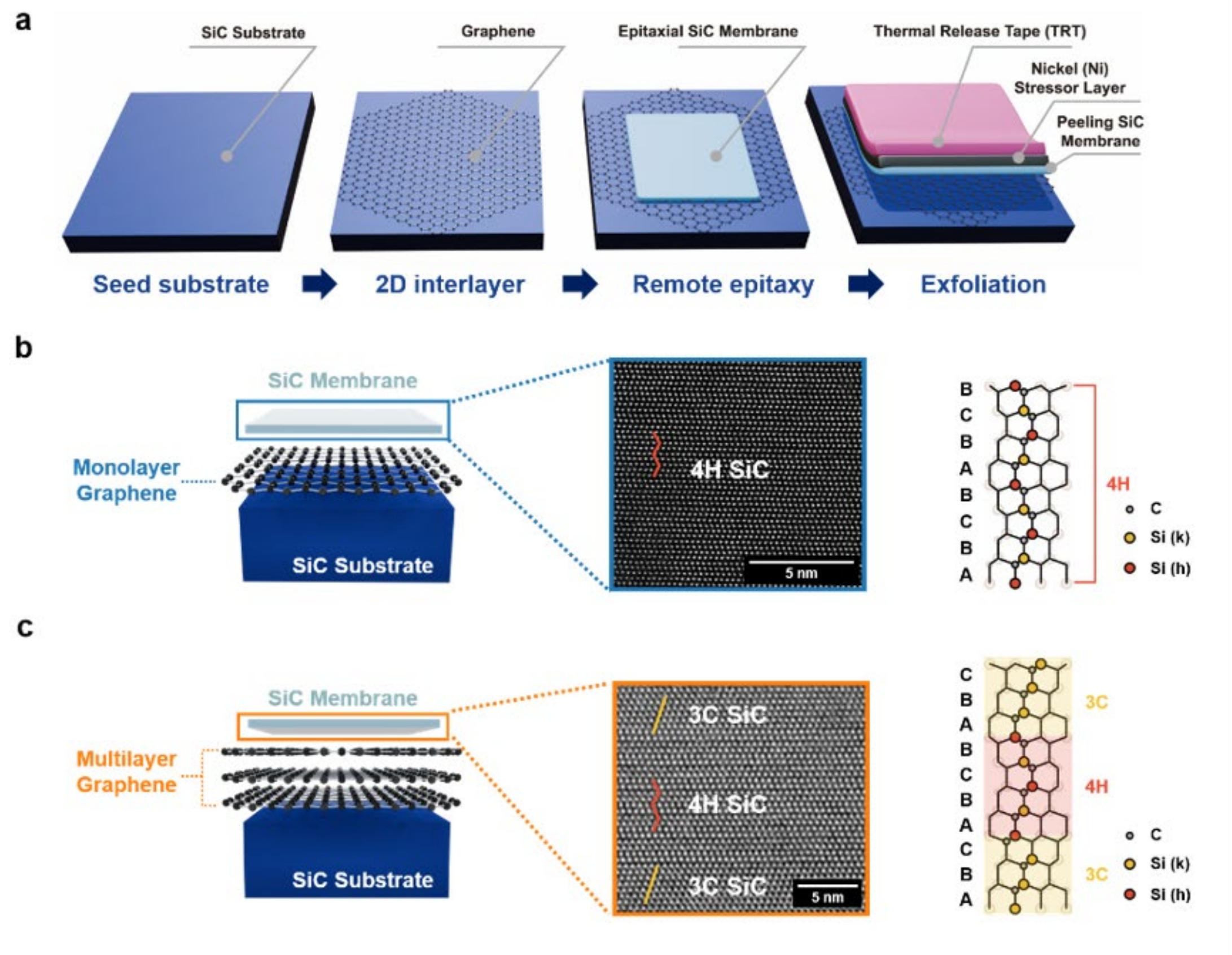

**Figure 1.** Remote epitaxy and structural characterization of phase-pure and heterophase SiC membranes. (a) Schematic illustration of the fabrication and mechanical exfoliation process of the epitaxial SiC membrane utilizing a graphene interlayer. (b) Schematic and cross-sectional TEM image of a phase-pure 4H-SiC membrane; the high-resolution TEM (HR-TEM) image confirms a highly crystalline and uniform atomic stacking configuration of the 4H polytype, alongside the corresponding atomic stacking model of the 4H-SiC structure. (c) Schematic and cross-sectional TEM image of the engineered heterophase membrane; the HR-TEM image explicitly reveals the alternating 3C/4H polytype interfaces, accompanied by a schematic illustration of the corresponding atomic stacking configuration, showing alternating 3C- and 4H-SiC regions across the heterophase interfaces.

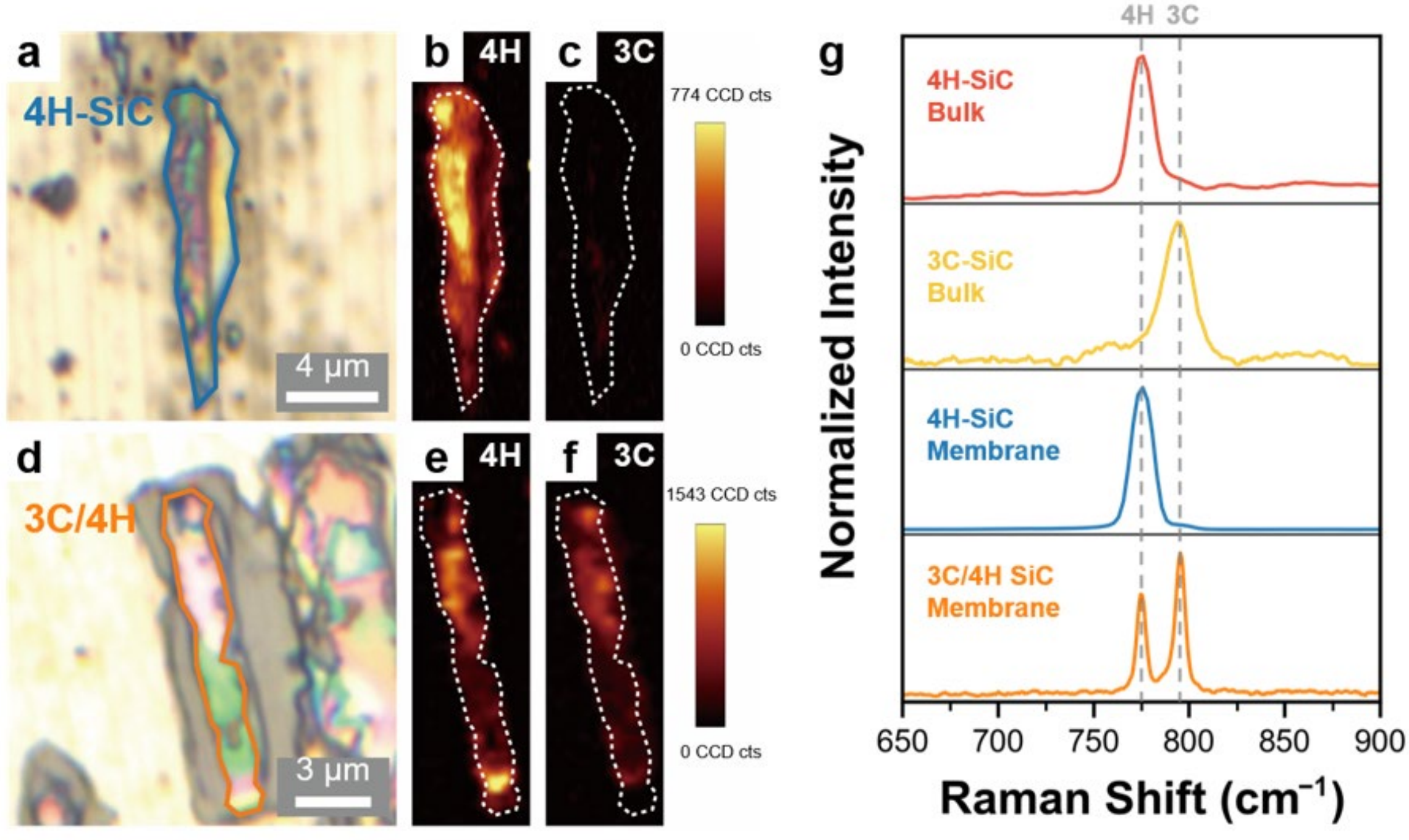

**Figure 2. Spatial and spectral Raman characterization of phase-pure and heterophase SiC membranes.** (a) Optical microscope image of the exfoliated phase-pure 4H-SiC membrane. (b,c) Confocal Raman intensity mappings of the phase-pure membrane, displaying a uniform 4H polytype distribution and an absence of the 3C phase signal. (d) Optical microscope image of the engineered heterophase membrane. (e, f) Confocal Raman intensity mappings of the heterophase membrane, revealing the spatial co-localization and distribution of both the 4H and 3C polytype domains. (g) Raman spectra of the exfoliated phase-pure 4H-SiC membrane (blue) and the engineered 3C/4H heterophase membrane (orange), compared with reference Raman spectra of bulk 4H-SiC (red) and bulk 3C-SiC (yellow), confirming their distinct transverse (TO) mode signatures at ~ 776 $cm^{-1}$and ~796 $cm^{-1}$, respectively.

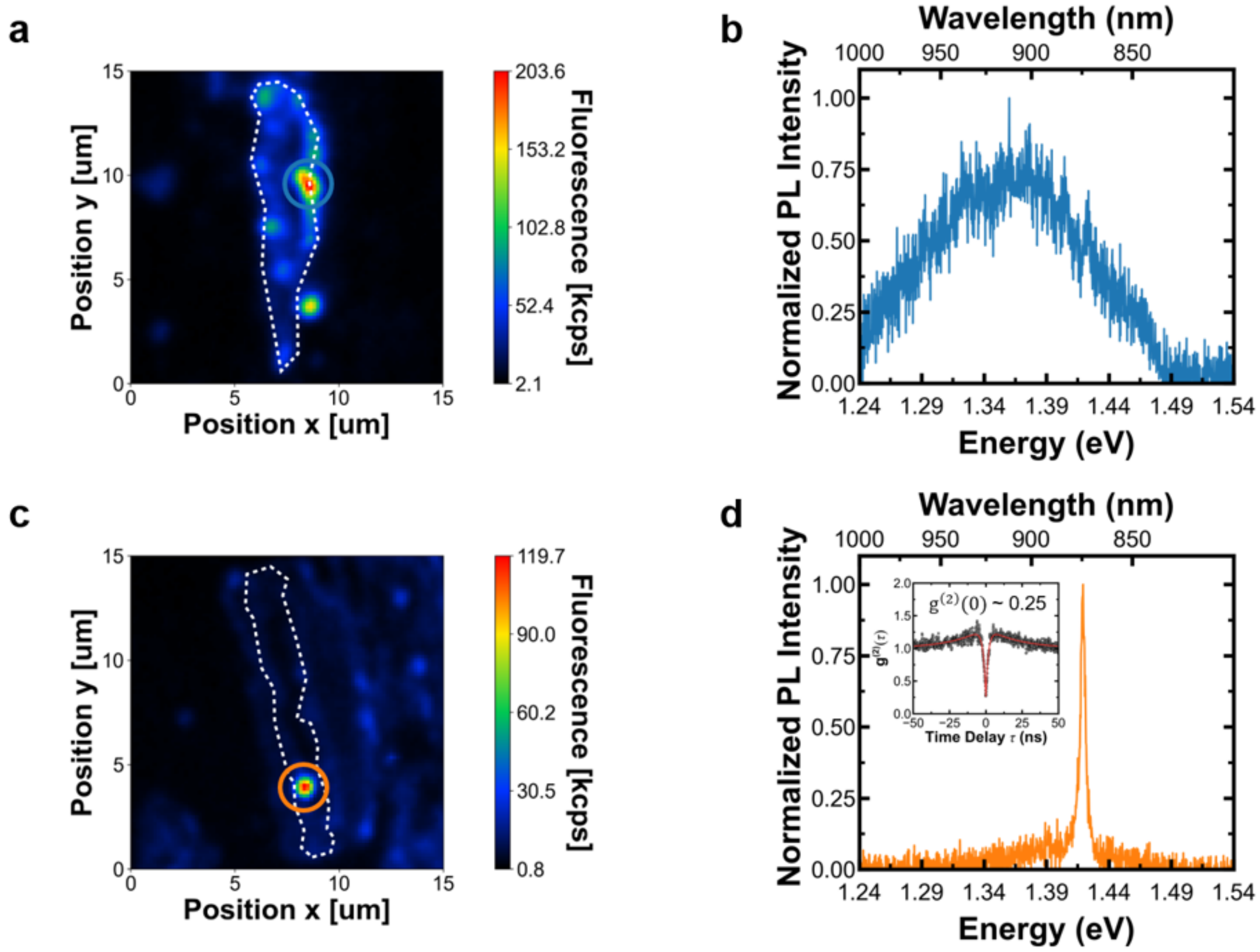

**Figure 3. Room temperature optical characterization of phase-pure and heterophase SiC membranes.** (a) Confocal PL mapping of the phase-pure 4H SiC membrane, with the membrane region indicated by the white dashed outline. The blue circle indicates a localized point defect emission corresponding to the PL spectrum in (b). (b) Room temperature PL spectrum of the phase-pure 4H SiC membrane. (c) Confocal mapping of the heterophase 3C/4H SiC membrane, with the membrane region indicated by the white dashed outline. The orange circle indicates a localized point defect emission corresponding to the PL spectrum in (d). (d) Room temperature PL spectrum of the heterophase 3C/4H SiC membrane showing narrow ZPL emission. *Inset:* Second-order intensity correlation function $g^{(2)}(0)$ of the single defect spot in the heterophase 3C/4H membrane.

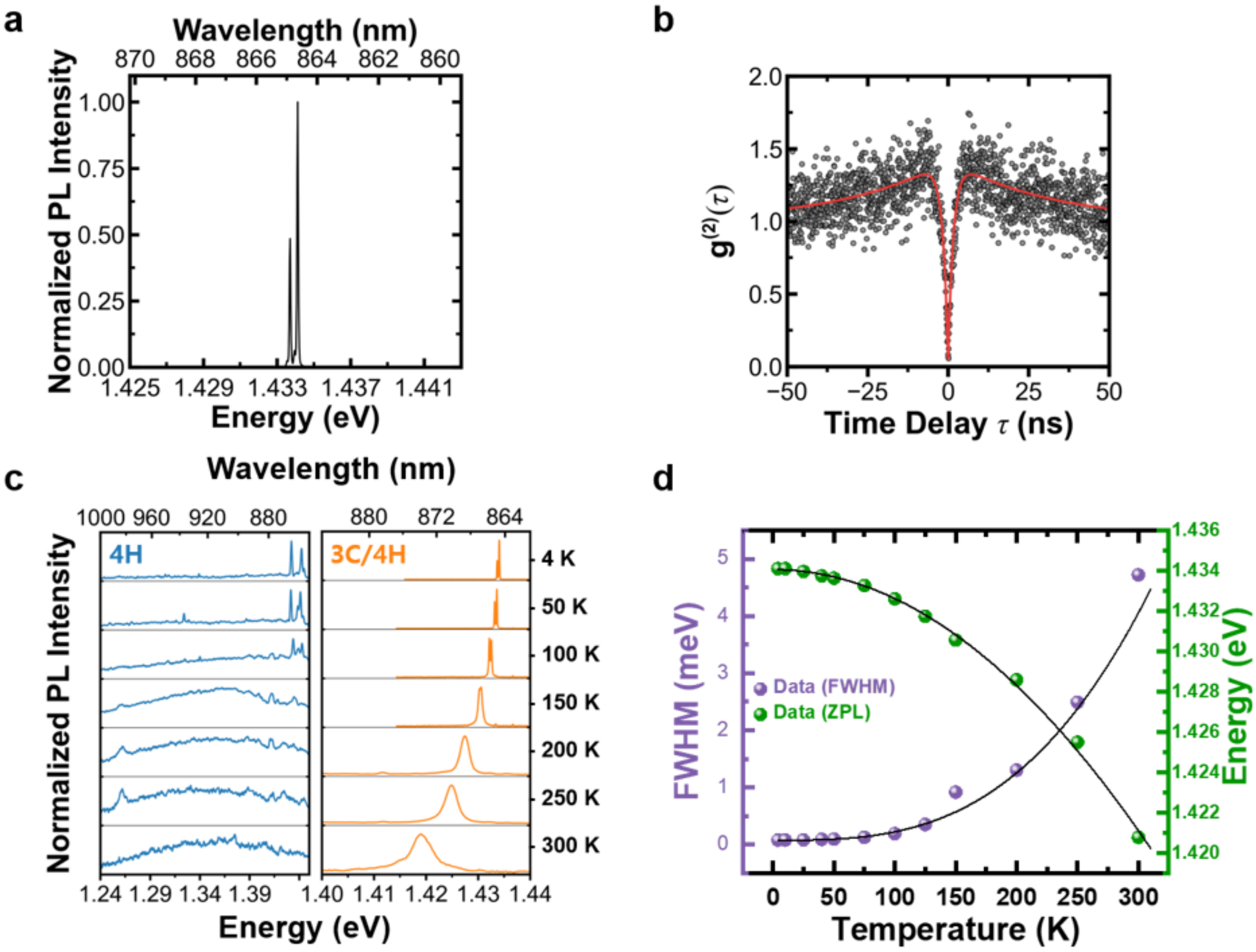

**Figure 4. Low-temperature and temperature-dependent PL characterization.** (a) High-resolution PL spectrum of the heterophase-associated defect identified in Figure 3c, d, measured at 4 K. (b) Second-order intensity correlation function $g^{(2)}(0)$ of the single defect measured at 4K. The red solid line represents the fit to the experimental data. (c) Temperature-dependent PL spectra from 4 K to 300 K comparing the phase-pure 4H-SiC membrane (blue) and the engineered 3C/4H heterophase SiC membrane (orange). (d) Full width at half maximum (FWHM, purple) and ZPL peak energy (green) of the single defect in the 3C/4H heterophase SiC membrane as a function of temperature. The black solid line indicates the theoretical fit for the line broadening.

# Supplementary Notes

### S1. Room-Temperature PL Spectra of Bulk and Pure-Phase Membrane Samples

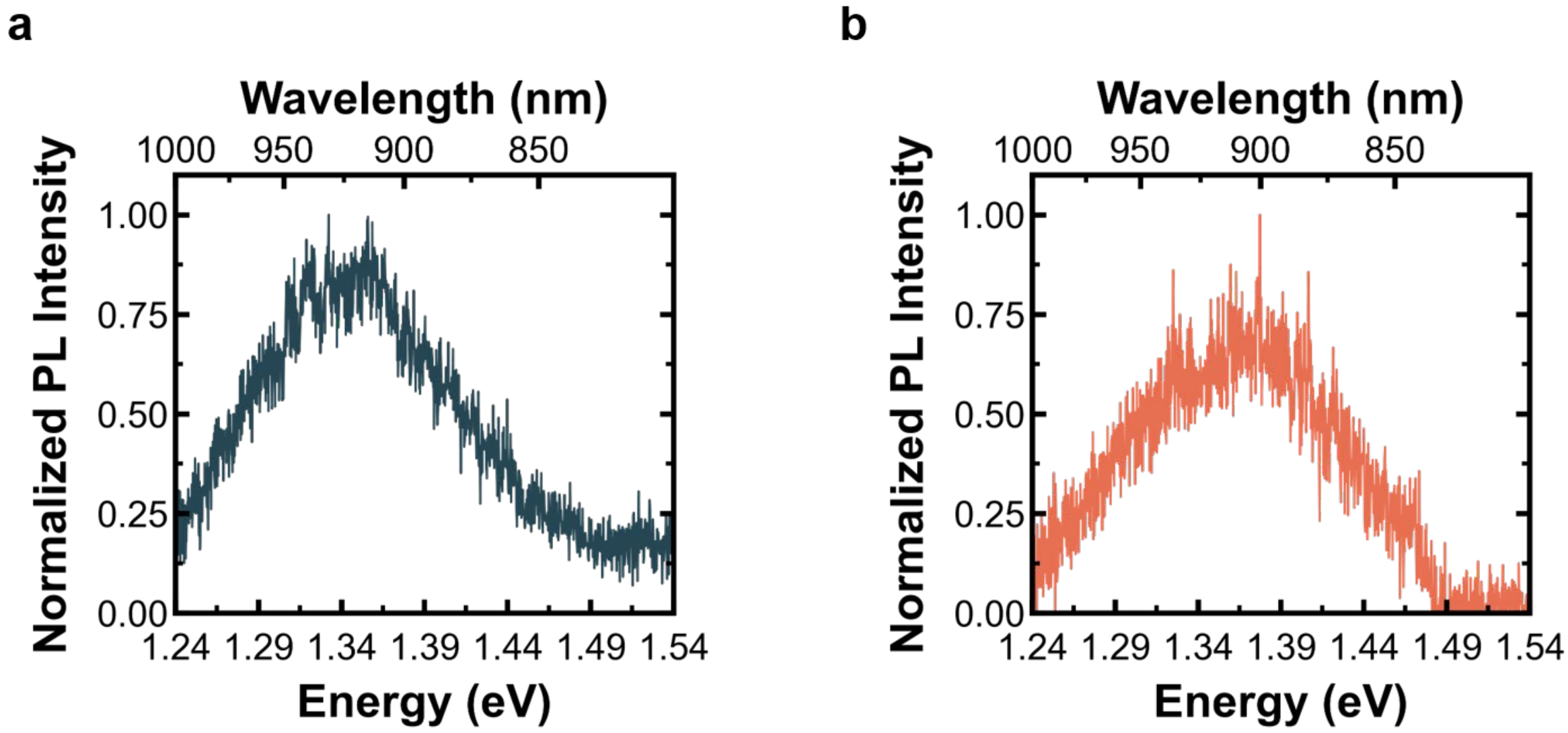

**Figure S1.** (a) Normalized RT PL spectrum measured from the bulk 4H-SiC substrate, showing a broad ensemble emission profile. (b) Normalized RT PL spectrum of a typical pure-phase transferred membrane, exhibiting similar phonon-broadened near-infrared emission without a clearly resolved zero-phonon line (ZPL).

**S2. Room-Temperature PL spectra across multiple membranes and power dependence of main-text SPE**

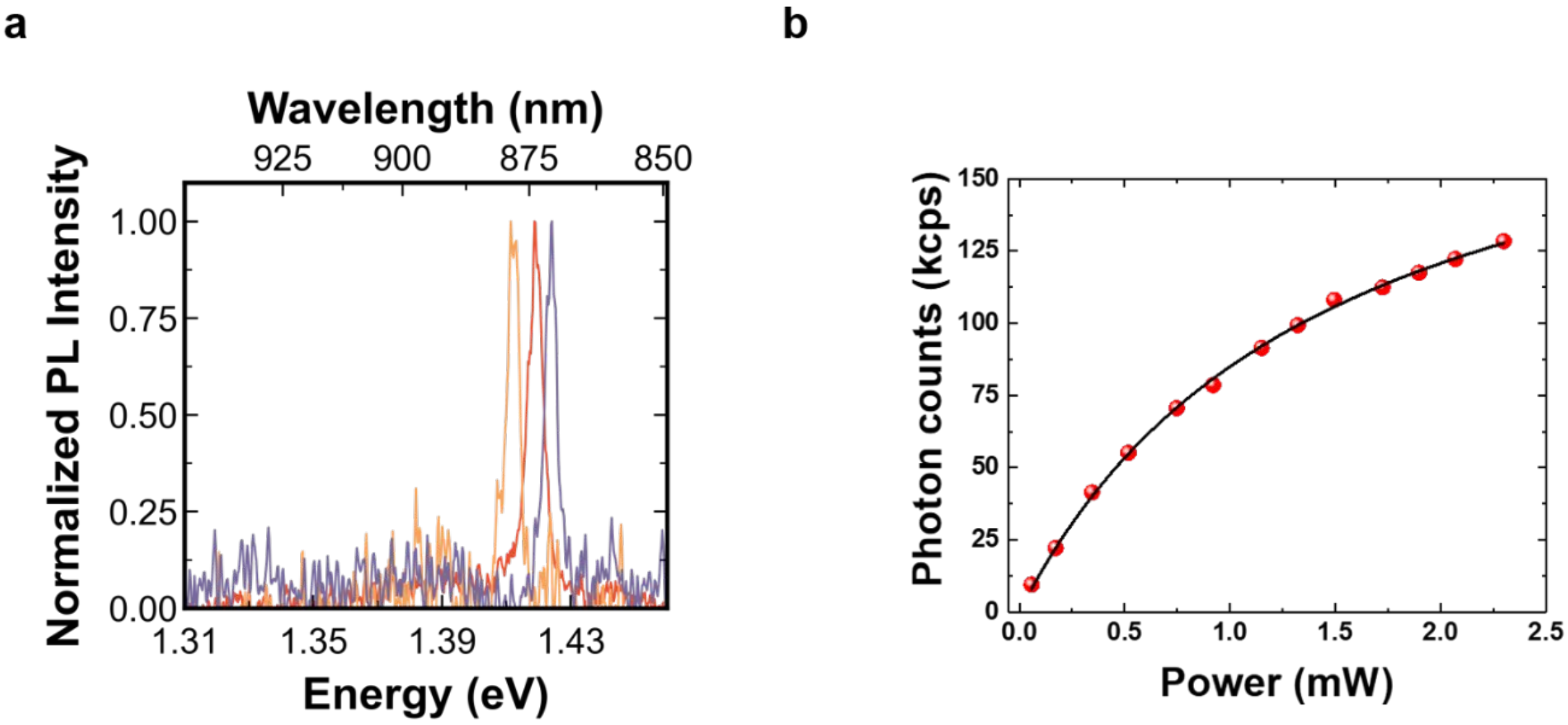

**Figure S2** (a) Normalized RT PL spectrum of single-photon emitters measured across different membranes, showing reproducible optical characteristics across transferred membranes b) Power-dependent PL intensity (saturation curve) recorded from SPE discussed in the main text, fitted using the standard two-level emitter saturation model, yielding a saturation intensity of $208 \pm 4$ kcounts/s and a saturation power ($P_{\mathrm{sat}}$) of 1.45 mW

## S3. Second-Order Correlation Measurements of Individual Emission Peaks

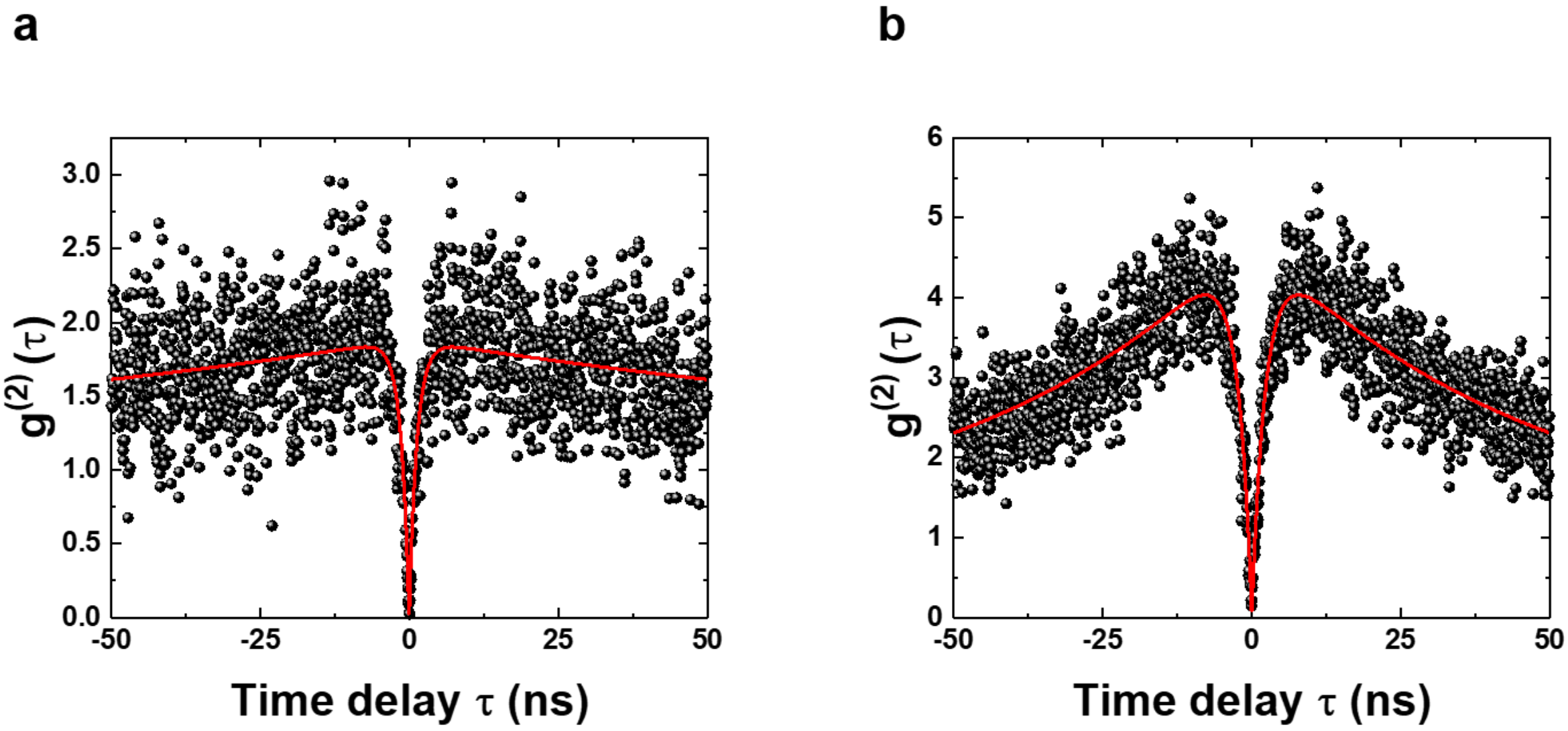

**Figure S3.** Low-temperature individual peak photon correlation statistics. Second-order autocorrelation functions $g^{(2)}(\tau)$ measured at 4 K for two distinct emission peaks isolated independently using a narrow fiber band-pass filter. Both individual measurements yield a raw, uncorrected minimum of $g^{(2)}(0) = 0.008$, confirming high-purity single-photon emission. The open black circles represent the raw photon coincidence data, and the solid red lines correspond to numerical fits utilizing three-level system. Variations in the bunching shoulder amplitudes highlight differing metastable state coupling dynamics between the two isolated transitions